\documentclass[aps,twocolumn,
preprintnumbers,showpacs,showkeys]{revtex4-1}
\usepackage{epsfig,dsfont}
\usepackage{amsmath}
\begin{document}  

\title{Dual representation for 1+1 dimensional fermions interacting with 3+1 dimensional U(1) gauge fields}
\author{Christof Gattringer}
\author{Vasily Sazonov}
\affiliation{Universit\"at Graz, Institut f\"ur Physik, Universit\"atsplatz 5, 8010 Graz, Austria}
 
\begin{abstract}
We study a system of nanowires, i.e., the theory of 1+1 dimensional massless fermions interacting with 3+1 dimensional U(1) gauge fields. 
When allowing for non-zero chemical potentials, this system has a complex action problem in the conventional formulation. 
We show that the partition sum can be mapped to a dual representation where the fermions correspond to dimers and oriented 
loops on 2-dimensional planes embedded in 4 dimensions. The dual degrees of freedom for the gauge fields are surfaces that either
are closed or bounded by the fermion loops. In terms of the dual variables the complex action problem is overcome and Monte Carlo 
simulations are possible for arbitrary chemical potentials.
\end{abstract}

\date{December 15, 2015}

\maketitle

\vskip20mm

\section{Introductory remarks}
Monte Carlo simulations of lattice field theories belong to the most powerful tools for the non-perturbative analysis of quantum field theories.
However, in the conventional formulation of a lattice field theory a Monte Carlo simulation is not always possible because the action $S$ 
can be complex
and the Boltzmann factor $e^{-S}$ does not have a probabilistic interpretation. This so-called complex action problem (or sign problem)
is typical for quantum field theories at finite densities. For overcoming the complex action problem a variety of approaches was
considered, such as re-weighting, complex Langevin dynamics or various series expansions. 
In some cases even a complete solution was achieved by
exactly rewriting the partition sum in terms of new degrees of freedom, so called dual variables (see, e.g., \cite{rev1,rev2,rev3,rev4,rev5,rev6}
for reviews).

The dual variables are loops for matter fields and surfaces for the gauge field degrees of freedom. The surfaces are either closed 
or bounded by matter flux. For fermions an additional problem appears: The loops pick up signs coming from the Grassmann nature
of the variables as well as from the Clifford algebra. This additional source of problems is the reason that there are only very few results
of a real and positive dualization of gauge theories with fermions.
 
In the current paper we generalize the successful dualization \cite{2dfermions,2dfermionsb} of massless fermions interacting with U(1) 
gauge fields in two dimensions to the case of 
QED with 3+1 dimensional U(1) gauge fields and a set of 1+1 dimensional fermions which may be interpreted as nano-wires. 
The fermions live on two-dimensional coordinate planes embedded in 3+1 dimensions.  At non-zero chemical potential the model exhibits a
complex action problem in the conventional representation which we solve by mapping the system to a dual representation. Our dual representation
for the set of nanowires also may be viewed as a partial dualization of full QED in four dimensions, since the fermion loops we consider here
are a subset of the loops appearing in full QED.

The model considered here is not only interesting as a further step in the program of finding dual representations for lattice field theories, 
but is directly related to the physics of graphene \cite{Castro2009}. More specifically, our model of QED$_4$ with fermion sections
corresponds to an array of graphene nano-wires. The peculiar properties of electrons in graphene, such as their Dirac spectrum
and their strong coupling to the electro-magnetic field opens the possibility to use such wires (nano-ribbons) as systems
for the construction of antennas sensitive to infrared radiation \cite{Tamagnone2012, Llatser2012, Yan2013}, and the model studied in 
our paper is the lattice regularization of such a system. 

\section{Definition of the model}
Our model of nanowires is defined as follows: For the electromagnetic field we use the conventional compact formulation, i.e., the gauge 
degrees of freedom are the link variables $U_\nu(n) \in$ U(1), $\nu = 1, \, ... \, 4$, 
living on the links of a 4-dimensional lattice. The sites of the lattice are denoted by $n$, the lattice size is $N_S^3 \times N_T$ and  
the gauge fields obey periodic boundary conditions. For the gauge action we use the Wilson form,
\begin{equation}
S_G[U] \; = \; - \beta \sum_{n} \sum_{\mu < \nu} \mbox{Re} \, U_{\mu \nu}(n) \; , 
\label{gaugeaction} 
\end{equation}
where $U_{\mu \nu}(n)$ denotes the plaquettes, 
\begin{equation}
U_{\mu \nu}(n) \; = \;
U_\mu(n) U_\nu(n+\hat{\mu}) U_\mu(n+\hat{\nu})^\star U_\nu(n)^\star \; .
\label{plaquettes}
\end{equation}

The fermions in our system are restricted to 1-dimensional spatial wires, which we choose to be straight lines parallel to one of 
the coordinate axes. The wires are labelled by an index $j = 1,2 \, ... \, N_w$, where $N_w$ is the total number of wires. Each 
wire is specified by a spatial direction and by the values of the two other spatial coordinates that are held fixed. Thus, together
with the euclidean time coordinate (the coordinate $\nu = 4$), a wire gives rise to a 2-dimensional space-time plane embedded 
in 4 euclidean dimensions. For example if the wire number $j$ is parallel to the 2-direction, then the corresponding plane is given by
\begin{equation}
p^{(j)} \!=\! \{n | n_1\! =\! n_1^{(j)}\!, n_2\! =\!1\, ..\,  N_S,  n_3\! =\! n_3^{(j)}\! , n_4 \!=\! 1\, ..\,  N_T\},
\end{equation}
where the coordinates $n_1^{(j)}$ and $n_3^{(j)}$ are held fixed -- they specify the spatial position of the wire axis which is parallel
to the 2-direction. For later use we also introduce functions $\theta^{(j)}_\nu(n)$ which are 1 for all links in the plane $p^{(j)}$
and 0 otherwise, i.e.,
\begin{equation}
\theta^{(j)}_\nu(n) \; = \; \left\{
\begin{array}{l}
1 \; \mbox{if} \; n \in p^{(j)} \; \mbox{and} \; n + \hat{\nu} \in p^{(j)} \\
0 \; \mbox{otherwise}
\end{array}
\right. 
\end{equation}

The fermions are restricted to the wires. Thus for each wire $j$ we have a set of Grassmann numbers
\begin{equation}
\overline{\psi}_j(n), \psi_j(n) \;\; \; \mbox{with} \;\; \; n \in p^{(j)} \; .
\end{equation}
We here use staggered fermions, such that the Grassmann variables do not carry a spinor index. They obey periodic 
boundary conditions in their spatial direction and anti-periodic boundary conditions in time.

The action $S_F[\psi, \overline{\psi},U]$ for our system of massless fermions on the wires interacting with the gauge field is given as
a sum over the contributions $S_F^{(j)}[\psi_j, \overline{\psi}_j,U]$ from the individual wires,
\begin{eqnarray}
&&S_F[\psi, \overline{\psi},U]  =  \sum_{j=1}^{N_w}  S_F^{(j)}[\psi_j, \overline{\psi}_j,U] \; ,
\label{factionsum}
\\
&& S_F^{(j)}[\psi_j, \overline{\psi}_j,U] = \sum_n \sum_{\nu=1}^4  \theta^{(j)}_\nu(n) \; \gamma_\nu(n) 
\times
\label{factionj}  \\
&&\qquad \qquad \frac{1}{2}\Big[ e^{\mu_j \delta_{\nu,4}}  \, 
\overline{\psi}_j(n) \, U_\nu(n) \, \psi_j(n\! +\!\hat{\nu}) \, - \, 
\nonumber \\
&& \qquad \qquad \qquad \qquad
e^{- \mu_j \delta_{\nu,4}} \, \overline{\psi}_j(n\!+\! \hat{\nu}) \, U_\nu(n)^\star \, \psi_j(n) \Big] .
\nonumber 
\end{eqnarray} 
By $\gamma_\nu(n)$ we denote the staggered sign factors,
$\gamma_1(n) = 1$, $\gamma_2(n) = (-1)^{n_1}$, $\gamma_3(n) = (-1)^{n_1 + n_2}$,
$\gamma_4(n) = (-1)^{n_1 + n_2 + n_3}$. On each wire $j$ we allow for a chemical potential $\mu_j$,
which couples to the temporal hopping terms in the canonical way.

The partition sum is given by
\begin{equation}
Z \; = \; \int D[U] D[\psi,\overline{\psi}\,] \, e^{-S_G[U] - S_F[\psi, \overline{\psi},U]} \; .
\label{partsum1}
\end{equation}
For integrating the gauge fields in the path integral we use a product of U(1) Haar measures $dU_\nu(n)$,
\begin{equation}
\int D[U] \; = \; \prod_{n,\nu} \int_{U(1)} dU_\nu(n) \; .
\end{equation}
The path integral measure for the fermions is a product of Grassmann measures for all
fermion degrees of freedom, 
\begin{eqnarray}
\int \! D[\psi,\overline{\psi}\,]  & \, = \, & \prod_{j=1}^{N_w} \; \int\! D[\psi_j,\overline{\psi}_j] \; ,
\label{grassmannfactorize} \\
\int \! D[\psi_j,\overline{\psi}_j]  & \, = \, &  \prod_n \int \! d \psi_j(n)  \,d \overline{\psi}_j(n) \; ,
\label{grassmanmeasure}
\end{eqnarray}
where for later use we have written the overall fermion measure as a product over the measures $\int \! D[\psi_j,\overline{\psi}_j]$ 
for the individual wires.

\section{Dual representation for individual wires}

The mapping of the partition sum (\ref{partsum1}) to its dual representation starts with observing that the
fermionic part $Z_F[U]$ of the partition sum can be factorized  into a product of the fermionic partition sums $Z_F^{(j)}[U]$ 
for the individual wires,
\begin{equation}
Z_F[U] \; = \;  \int \! D[\psi,\overline{\psi}\,] \; e^{- S_F[\psi, \overline{\psi},U]}  \; = \; \prod_{j=1}^{N_w} Z_F^{(j)}[U] \; ,
\label{zfactorization}
\end{equation}
where the fermionic partition sum $Z_F^{(j)}[U]$ for wire $j$ is given by
\begin{equation}
Z_F^{(j)} [U] \; = \;  \int \! D[\psi_j,\overline{\psi}_j] \; e^{- S_F^{(j)}[\psi_j, \overline{\psi}_j,U]} \; .
\label{2dfermions}
\end{equation}
The factorization (\ref{zfactorization}) follows from the facts that the fermion action is a sum over 
the contributions from the wires (\ref{factionsum}) and the factorization of the Grassmann integration 
measure (\ref{grassmannfactorize}).

\begin{figure*}[t]
\hspace*{3mm}
\includegraphics[height=7.2cm,type=pdf,ext=.pdf,read=.pdf]{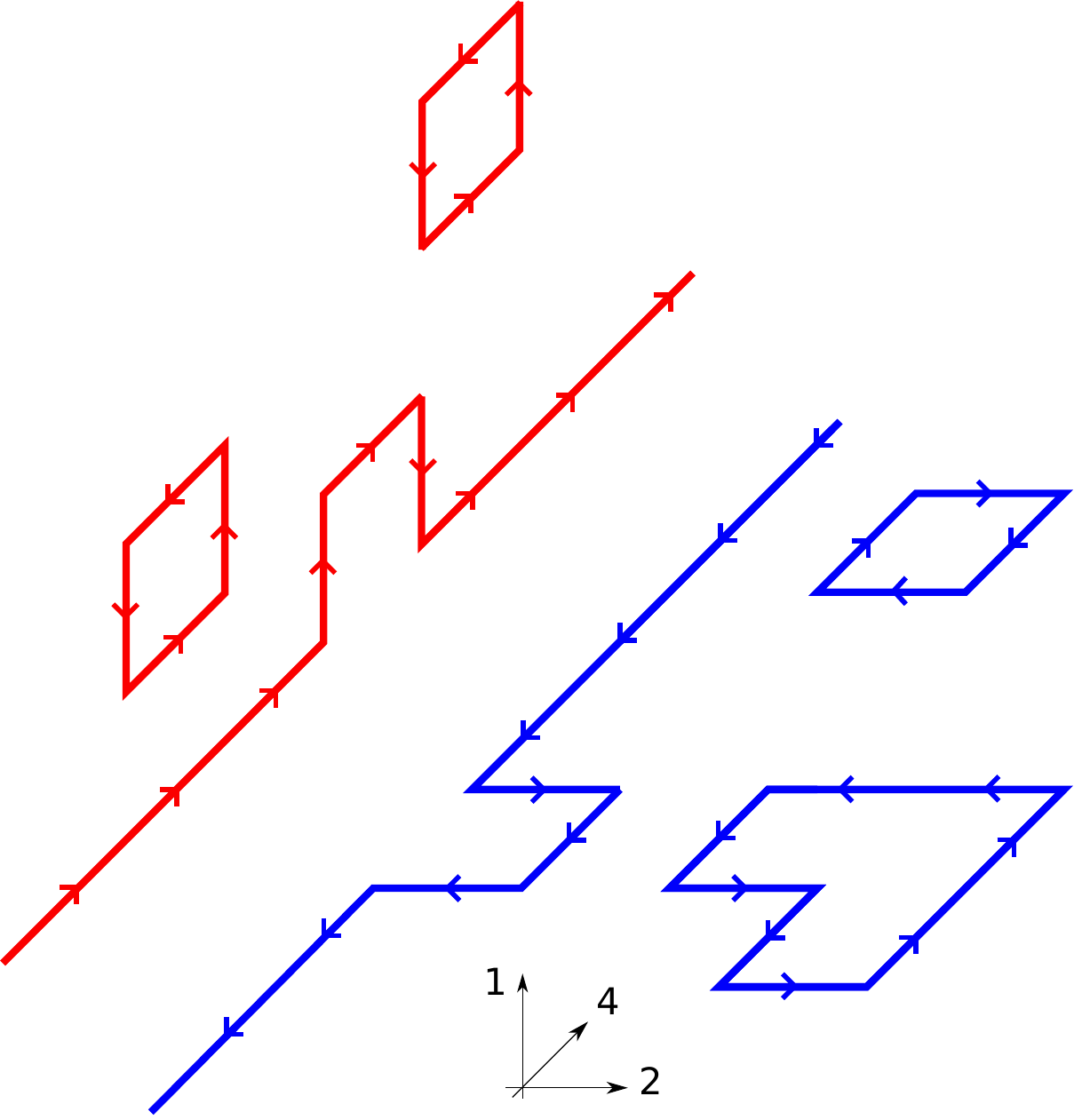}
\hspace{19mm}
\includegraphics[height=7.2cm,type=pdf,ext=.pdf,read=.pdf]{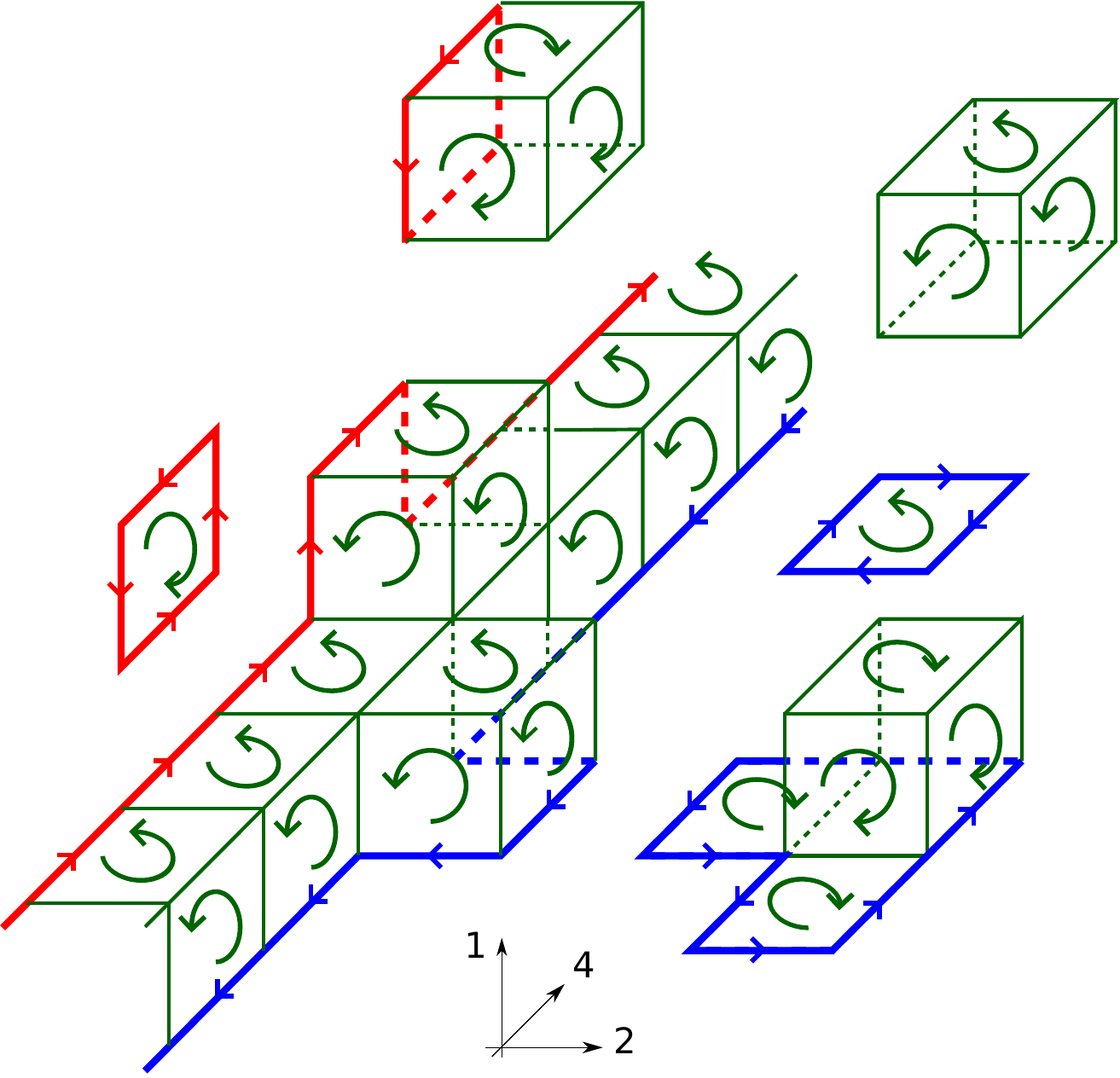}
\caption{Fermion paths and their saturation with plaquettes. In the lhs.\ plot we show fermion loops in the 1-4 and the 2-4 planes (note that in each 
plane one of the loops closes periodically around the compact time direction (= 4-direction). The remaining sites in both planes are filled with dimers
which are not shown here. In the rhs.\ plot we show a possible configuration of the 
plaquette occupation numbers, such that at each link the total flux from fermion loops and occupied plaquettes vanishes. Note that this plot also shows a pure gauge contribution -- the cube in the upper right corner.}
\label{fig1}
\end{figure*}

The key observation for obtaining the real and positive dual representation is to note that the 
partition sums $Z_F^{(j)} [U]$ are identical to the partition sum for 2-dimensional massless staggered fermions  
in a U(1) background gauge field. It is obvious that indeed the fermions in the partition sum $Z_F^{(j)} [U]$ 
with action given by (\ref{factionj}) can propagate only in a 2-dimensional plane. The only difference to the standard
formulation of 2-dimensional staggered fermions is that here the 4-dimensional staggered sign function is used. 

Let us for example consider a wire in the 3-direction. Then the fermions have hopping terms in the 3- and the 4-direction. 
The corresponding staggered sign functions thus are the spatial sign function $\gamma_3(n) = (-1)^{n_1+n_2}$ and the temporal one,
$\gamma_4(n) = (-1)^{n_1 + n_2 + n_3}$. In the standard 2-dimensional formulation one uses $\gamma_1(n) = 1$ for the spatial hops and
$\gamma_2(n)= (-1)^{n_1}$ for the temporal hops. However, when mapping the partition sum to a sum of loops this difference plays no role:
In both, the 4-d and the 2-d case integrating out the fermions gives rise to configurations of loops and dimers in a plane which is
the 1-2 plane for the 2-d case and the $k$-4 plane for a wire in direction $k = 1,2$ or 3. For the dimers the corresponding 
$\gamma_\nu(n)$ factor is squared and drops out. For a loop the product of the factors $\gamma_\nu(n)$ for all links of the loop 
have to be taken into account. For our example of a loop in the 3-4 plane the relevant staggered signs $\gamma_3(n)$ and $\gamma_4(n)$ both 
have a factor $(-1)^{n_1 + n_2}$ which is the same for all links in the loop, since $n_1$ and $n_2$ only define the position of the 3-4 plane 
carrying the loops. Thus for a given loop $l$ we have an overall factor $(-1)^{(n_1+n_2)*|l|}$ where $|l|$ denotes the length of the loop. Since 
all loops have even length (note that $N_S$ and $N_T$ are even for staggered fermions) this factor drops out. Thus the sign factors of 
the loops on planes embedded in 4 dimensions are the same as in the genuine 2-d case. These signs are related to shape properties
of the loops such as the overall length, the number of plaquettes inside a loop and the total winding number of the loops around the 
compactified time direction. For massless staggered fermions in 2 dimensions it was shown \cite{2dfermions,2dfermionsb} that all 
signs cancel when the partition function is represented as a sum of loop and dimer configurations. Since this result depends only on shape properties
of the loops, it carries over to the loops on planes in 4-d. 

Strictly speaking this result holds only after the gauge fields
are integrated out, a step we discuss in detail below. 
This enforces overall charge neutrality due to Gauss' law, which in turn gives rise to equal temporal winding numbers of loops for 
positive and negative charge. Thus the overall temporal winding number is even and the sign contribution from the anti-periodic temporal boundary conditions is $+1$. We remark at this point that, as in the 2-d case, a mass term spoils the positivity due to the appearance of monomer terms \cite{2dfermions}.

We obtain the following exact representation for the fermionic partition sum $Z_F^{(j)} [U]$ of the wire $j$ 
(we have dropped an overall irrelevant factor $2^{-N_S N_T}$):
\begin{equation}
Z_F^{(j)} [U] = \!\!  \sum_{\{l^{(j)},d^{(j)}\!\}}  \!\! e^{\mu_j N_T W[l^{(j)}]} \, \prod_{l^{(j)}} \prod_{(n,\nu)\in l^{(j)}} U_\nu(n)^{s_\nu(n)} .
\label{zfjdual}
\end{equation}
The sum runs over all configurations $\{l^{(j)},d^{(j)}\!\}$ of oriented loops $l^{(j)}$ and dimers $d^{(j)}$ in the plane $p^{(j)}$ relevant for wire $j$. 
The configurations have to be such that each site in $p^{(j)}$ is either the endpoint of a dimer or run through by a loop. All loops $l^{(j)}$ 
are dressed with the gauge links $U_\nu(n)$ along their contour where the exponent $s_\nu(n)$ takes care of the orientation the link 
$(n,\nu)$ is run through by the loop, i.e.,  $s_\nu(n) = +1$ for links that are run through in positive direction and $s_\nu(n) = -1$ for negative
direction. We refer to the $s_\nu(n)$ as the link occupation numbers.
Finally, the chemical potential is coupled to the total net winding number $W[l^{(j)}]$ of all loops around compactified time. Here 
$N_T$ is the temporal extent of the lattice, which coincides with the inverse temperature $\beta$ in lattice units, i.e., $N_T = \beta$. Since
in the grand canonical ensemble the chemical potential couples to the particle number ${\cal N}$ in the form
$e^{\mu \beta {\cal N}}$, we can identify the total net winding number $W[l^{(j)}]$ as the number of charges on wire $j$.

\section{Integrating out the gauge fields}

The total fermionic partition sum $Z_F[U]$ is a product over the partition sums (\ref{zfjdual}) for the individual wires. These are sums over loops 
which are dressed with the gauge fields $U_\nu(n)$. For illustration, in the lhs.\ plot of Fig.~\ref{fig1} we show a configuration of loops in two 
planes. Note that the remaining sites in both planes are filled with dimers which we do not show to avoid overcrowding the plot.

To obtain the full partition 
function $Z$ the fermionic partition function $Z_F[U] = \prod_{j=1}^{N_w} Z_F^{(j)} [U]$ 
still has to be integrated over the gauge fields with the Boltzmann factor $e^{-S_G[U]}$ for the gauge fields,
\begin{eqnarray}
Z & = & \int \!\! D[U] \; e^{-S_G[U]} \, \prod_{j=1}^{N_w} Z_F^{(j)} [U] 
\\
&= & \left( \prod_{j=1}^{N_w} \sum_{\{l^{(j)},d^{(j)}\!\}} \right) \left( \prod_{j=1}^{N_w}  e^{\mu_j N_T W[l^{(j)}]} \right) \times
\nonumber \\
&& \quad \quad \int \!\! D[U] \; e^{-S_G[U]}  \, \prod_{j=1}^{N_w} \prod_{l^{(j)}} \prod_{(n,\nu)\in l^{(j)}} \! \!\big(U_\nu(n)\big)^{s_\nu(n)} .
\nonumber
\end{eqnarray}
For integrating the gauge fields we write the Boltzmann factor of the gauge fields as a product of individual factors \cite{U1,surfaceworm},
\begin{eqnarray}
e^{-S_G[U]} & = & \prod_{n, \mu < \nu} e^{\, \frac{\beta}{2} [ U_{\mu \nu}(n) + U_{\mu \nu}(n)^{-1} ]} 
 \label{gaugeseries} \\
 & = & \prod_{n, \mu < \nu}  \; \sum_{p_{\mu \nu} \in \mathds{Z}} I_{p_{\mu \nu}(n)} ( \beta ) \; \big( U_{\mu \nu}(n) \big)^{p_{\mu \nu}(n)} ,
\nonumber 
\end{eqnarray}
where in the second line we have expanded these individual factors, using the fact, that the corresponding exponential 
is the generating function of the modified Bessel functions $I_p$. Since the product runs over all plaquettes,
we have one expansion index $p_{\mu \nu}(n) \in \mathds{Z}$ for each plaquette. The $p_{\mu \nu}(n)$ are referred to as
plaquette occupation numbers and constitute the dual variables which describe the gauge degrees of freedom.  

In the form (\ref{gaugeseries}) the link variables in the Boltzmann factor $e^{-S_G[U]}$ are brought down from the exponent
and appear in powers $\big( U_{\mu \nu}(n) \big)^{p_{\mu \nu}(n)}$ of the corresponding plaquettes (\ref{plaquettes}).
For each link $(n,\nu)$ we can now collect the link variables $U_\nu(n)$ that are contributed from the plaquettes 
attached to that link and from all loops that run through the link $(n,\nu)$. Thus each link variable $U_\nu(n)$ appears in the form 
$\big( U_{\nu}(n) \big)^{a_{\nu}(n)}$, where the exponent $a_{\nu}(n)$ combines the link occupation numbers $s_\nu(n)$ of all 
loops running through the link $(n,\nu)$ and the plaquette occupation numbers $p_{\mu \nu}(n)$ of all plaquettes attached to that link. 

The integration of the gauge fields now is straightforward. The key formula is
\begin{equation}
\int D[U] \, \prod_{n,\nu} \big( U_{\nu}(n) \big)^{a_{\nu}(n)} \, = \, \prod_{n,\nu} \delta({a_{\nu}(n)}) \; ,
\end{equation}
which follows from the fact that integrating powers of a U(1) phase is a representation of the Kronecker delta, which here is denoted as
$\delta(k) \equiv \delta_{k,0}$. Integrating out the gauge fields thus gives rise to a product over constraints at all links: The total 
flux from the loops containing that link and the plaquettes attached to that link has to vanish. 

These constraints have a simple interpretation: Admissible plaquette occupation numbers are such that the plaquettes form 2-dimensional 
surfaces of equal occupation numbers. The surfaces can have boundaries where the flux on the links of the boundaries is compensated
by the fermion loops. Alternatively the plaquette occupation numbers can form closed surfaces. In the rhs.\ plot of Fig.~\ref{fig1} we show an example 
of an admissible configuration of plaquette occupation numbers for the configuration of loops shown in the lhs.\ plot. Non-vanishing plaquette 
numbers are indicated by arrows on the plaquettes (in this configuration only plaquette occupation numbers $\pm 1$ appear).
 
Thus we can summarize the dual representation as follows:
\begin{equation}
Z =  \left( \, \prod_{j=1}^{N_w} \sum_{\{l^{(j)},d^{(j)}\!\}} \!\!\!\! e^{\, \mu_j N_T W[l^{(j)}]} \!\right) \! \sum_{\{p\}}  
\prod_{n,\mu < \nu} I_{p_{\mu \nu}(n)}\!\left( \beta \right) \, .
\label{zfinal2}
\end{equation}
The partition sum is a sum over all admissible configurations $\{l^{(j)}, d^{(j)} \}$  of loops and dimers in all planes $j = 1,2, \, ... \, N_w$.
The admissible configurations are such that each site of the plane is either run through by a loop or is the endpoint
of a dimer. The chemical potentials $\mu_j$ for the wires enter via the total temporal winding number $W[l^{(j)}]$ of the loops. 
For each given loop configuration one has to sum over all configurations of the plaquette occupation numbers $p_{\mu \nu}(n)$  
such that the plaquettes form surfaces that are either closed or bounded by the loops. The corresponding Boltzmann factor is the 
product of the Bessel functions $I_{p_{\mu \nu}(n)}(\beta)$.

We stress that in the dual representation (\ref{zfinal2}) all contributions are real and positive, such that the complex action problem is solved. 
In terms of the dual variables a Monte Carlo simulation of the system of nanowires is possible for arbitrary values of the chemical potentials. 
Suitable powerful algorithms for U(1) gauge fields in the dual representation were developed in the context of abelian gauge Higgs systems
\cite{U1,surfaceworm} and can be adapted for simulating (\ref{zfinal2}) in a straightforward way. We are currently implementing such a simulation
starting with the simpler and less computer time consuming 2-dimensional case \cite{2dfermions,2dfermionsb}.
\relax

\section{Comments on dual observables}

Observables can be generated either by derivatives with respect to the couplings of the theory, i.e., $\beta$ and the $\mu_j$, or by derivatives with 
respect to suitably introduced sources. These derivatives can then be evaluated also in the dual representation of the partition sum and in this way
one identifies the dual form of the observables.

For the example of the plaquette expectation value we find ($V = N_S^3 N_T$)
\begin{eqnarray}
\langle U_p \rangle & \, = \, & \frac{1}{6 V}  \sum_{n,\mu < \nu}  \mbox{Re} \, \langle U_{\mu \nu}(n) \rangle 
\; = \; \frac{1}{6 V}  \frac{\partial}{\partial \beta} \ln Z 
\nonumber \\
& = & \frac{1}{6 V}  \sum_{n,\mu < \nu}  \, \left\langle \frac{I_{p_{\mu \nu}(n)}\!\left( \beta \right)^\prime}{I_{p_{\mu \nu}(n)}\!\left( \beta \right)}
\right\rangle 
\; ,
\end{eqnarray}
where the expectation value on the rhs.~is understood in terms of the dual variables. 
$I_p(\beta)^\prime$ denotes the derivative of the generalized Bessel function with respect to $\beta$. 
In a similarly simple way we obtain the particle number density $n^{(j)}$ in wire $j$ as a derivative with respect to $\mu_j$,
\begin{equation}
\langle n^{(j)} \rangle \; = \; \frac{1}{N_S N_T}  \frac{\partial}{\partial \mu_j} \ln Z 
\; = \; \frac{1}{N_S}   \left\langle W[l^{(j)}] \right\rangle \; .
\end{equation}
The corresponding susceptibilities are obtained by another derivative with respect to $\beta$ and $\mu_j$, respectively, and one finds
that the dual representations of bulk observables are moments of the dual variables and their weights. 

However, also $n$-point functions can be represented in the dual form. One can couple real valued source terms $b^{(j)}_\nu(n)$, 
$\overline{b}^{(j)}_\nu(n)$ to the fermionic nearest neighbor terms such that they enter the fermion action in the form 
(we here leave out the staggered factors $\gamma_\nu(n)$ and the chemical potential terms),
\begin{eqnarray}
&& b^{(j)}_\nu(n) \; \overline{\psi}_j(n) U_\nu(n) \psi_j(n+\hat{\nu}) \; , \\
&& \overline{b}^{(j)}_\nu(n)  \; \overline{\psi}_j(n+\hat{\nu}) U_\nu(n)^\star \psi_j(n)  \; .
\nonumber 
\end{eqnarray}   
The source terms couple to the gauge invariant nearest neighbor terms. They can be used to obtain $n$-point functions of various currents by
suitable derivatives with respect to $b^{(j)}_\nu(n)$ and $\overline{b}^{(j)}_\nu(n)$, and subsequent replacement 
$b^{(j)}_\nu(n) \rightarrow 1$ and $\overline{b}^{(j)}_\nu(n) \rightarrow 1$. 

When dualizing the fermions the sources appear in the same
way as the gauge links, i.e., they appear as factors along the loops. The subsequent integration of the gauge fields then can be 
done as before. The resulting dual partition function with sources has essentially the same form as before, but the links of the loops are now dressed 
with the sources: $b^{(j)}_\nu(n)$ for links that are run through in positive direction and $\overline{b}^{(j)}_\nu(n)$ for negative orientation. Now
taking the derivatives is straightforward and $n$-point functions of $\overline{\psi}_j(n) U_\nu(n) \psi_j(n+\hat{\nu})$ and 
$\overline{\psi}_j(n+\hat{\nu}) U_\nu(n)^\star \psi_j(n)$ in the dual representation turn into 
the correlators of the corresponding links. More explicitly, the $n$-point function
receives contributions from all those dual configurations where all $n$ links are occupied by a loop with the correct orientation. 

\section{Concluding remarks}
In this paper we present a new example of a successful solution of a complex action problem by an exact mapping of the partition function 
to dual variables. The model analyzed is a system of nano-wires interacting with the electromagnetic field. The 
solution makes use of a previous result for 1+1 dimensional massless fermions, where a loop representation was found in which all minus signs 
from the fermionic nature were eliminated. The coupling to the gauge fields can easily be generalized to the 4-dimensional case, and 
several 1+1 dimensional fermions can be combined into a a system of nanowires. The resulting dual form of the partition function is
a sum over 2-d loops for the fermions in the wires and surfaces for the gauge fields. The surfaces are either closed or bounded by the
fermion loops. 

Arbitrary chemical potentials $\mu_j$ can be coupled for the different nanowires and the corresponding complex action problem of the conventional 
representation is completely overcome in the dual representation, i.e., all weights in the dual representation are real and positive for
arbitrary $\mu_j$. The chemical potentials couple to the temporal net winding numbers of the loops and give different weight to forward 
and backward winding, revealing the interpretation of the temporal winding number as the dual form of the net particle number.  

We remark that the results presented here correspond to a partial dualization of full QED, since the 1+1 dimensional loops appearing 
here are subsets of the full 4-dimensional loops of QED. A different partial real and positive dualization of QED based on hopping
expansion techniques was presented in \cite{QED4}. We expect that the program of finding dual representations for 
systems of interacting fermions will profit from analyzing various sub systems, and we hope that the results presented here will contribute to 
finding more general dual forms. 

\vspace{5mm}
\noindent
{\bf Acknowledgements:} We thank Daniel G\"oschl and Thomas Kloiber for interesting discussions. This work is supported by the Austrian 
Science Fund FWF, through the DK {\sl Hadrons in Vacuum, Nuclei, and Stars} (FWF DK W1203-N16) and by FWF Grant I 1452-N27. 
We also acknowledge partial support from DFG TR55, {\sl ``Hadron Properties from Lattice QCD''}.

\newpage
\bibliographystyle{apsrev}

\begin{thebibliography}{17}

\bibitem{rev1}
  D.~Sexty,
  {\sl New algorithms for finite density QCD,}
  PoS LATTICE {\bf 2014} (2015) 016 [arXiv:1410.8813].

\bibitem{rev2}
  C.~Gattringer,
  {\sl New developments for dual methods in lattice field theory at non-zero density,}
  PoS LATTICE {\bf 2013} (2014) 002
  [arXiv:1401.7788].
  
 \bibitem{rev3}
  G.~Aarts,
  {\sl Complex Langevin dynamics and other approaches at finite chemical potential,}
  PoS LATTICE {\bf 2012} (2012) 017
  [arXiv:1302.3028].

\bibitem{rev4}
  U.~Wolff,
  {\sl Strong coupling expansion Monte Carlo,}
  PoS LATTICE {\bf 2010} (2010) 020
  [arXiv:1009.0657].

\bibitem{rev5}
  P.~de Forcrand,
  {\sl Simulating QCD at finite density,}
  PoS LAT {\bf 2009} (2009) 010
  [arXiv:1005.0539].

\bibitem{rev6}
  S.~Chandrasekharan,
  {\sl A New computational approach to lattice quantum field theories,}
  PoS LATTICE {\bf 2008} (2008) 003
  [arXiv:0810.2419].

\bibitem{2dfermions}
  C.~Gattringer, T.~Kloiber and V.~Sazonov,
  {\sl Solving the sign problems of the massless lattice Schwinger model with a dual formulation,}
  Nucl.\ Phys.\ B {\bf 897} (2015) 732
  [arXiv:1502.05479].
 
\bibitem{2dfermionsb}
  C.~Gattringer, T.~Kloiber and V.~Sazonov,
  {\sl Dual representation for massless fermions with chemical potential and U(1) gauge fields,}
  PoS LATTICE {\bf 2015} (2015) 195
  
\bibitem{Castro2009}
  A.H.~Castro Neto, F.~Guinea, N.M.R.~Peres, K.S.~Novoselov and A.K.~Geim,
  {\sl The electronic properties of graphene},
  Rev.~Mod.~Phys.~{\bf 81} (2009) 109.
 
\bibitem{Tamagnone2012}
   M.~Tamagnone, J.S.~Gomez-Di­az, J.R.~Mosig and J.~Perruisseau-Carrier,
   {\sl Reconfigurable terahertz plasmonic antenna concept using a graphene stack},
   Appl.~Phys.~Lett.~{\bf 101} (2012) 214102.
 
\bibitem{Llatser2012}
I.~Llatser, C.~Kremers, A.~Cabellos-Aparicio, J.~Miquel Jornet, E.~Alarcon and D.N.~Chigrin,
    {\sl Graphene-based nano-patch antenna for terahertz radiation},
     Photonics and Nanostructures - Fundamentals and Applications {\bf 10} (2012) 353.
      
\bibitem{Yan2013}
   H.~Yan, T.~Low, W.~Zhu, Y.~Wu, M.~Freitag, X.~Li, F.~Guinea, P.~Avouris and F.~Xia,
   {\sl Damping pathways of mid-infrared plasmons in graphene nanostructures},
   Nature Photonics {\bf 7} (2013) 394.  

\bibitem{U1}
  Y.D.~Mercado, C.~Gattringer and A.~Schmidt,
  {\sl Dual Lattice Simulation of the Abelian Gauge-Higgs Model at Finite Density: An Exploratory Proof of Concept Study,}
  Phys.\ Rev.\ Lett.\  {\bf 111} (2013) 14,  141601
  [arXiv:1307.6120].
  
\bibitem{surfaceworm}
  Y.D.~Mercado, C.~Gattringer and A.~Schmidt,
  {\sl Surface worm algorithm for abelian Gauge-Higgs systems on the lattice,}
  Comput.\ Phys.\ Commun.\  {\bf 184} (2013) 1535
  [arXiv:1211.3436].

\bibitem{QED4}
  M.~Kniely and C.~Gattringer,
  {\sl Dual simulation of finite density lattice QED at large mass,}
  PoS LATTICE {\bf 2014} (2015) 206
  [arXiv:1502.00788].

\end{thebibliography}

\end{document}